# Relation between resistance drift and optical gap in phase change materials

J C Martinez* and R E Simpson
Engineering and Production Development
Singapore University of Technology and Design
8 Somapah Rd, Singapore 487372

The optical contrast in a phase change material is concomitant with its structural transition. We connect these two by first recognizing that Friedel oscillations couple electrons propagating in opposite directions and supply an additional Coulomb energy. As the crystal switches phase, this energy acquires time dependence and the Landau-Zener mechanism operates, steering population transfer from the valence to the conduction band and vice versa. Spectroscopy suggests that the oscillator energy $E_0$ dominates the optical properties and a calculation involving the crystalline field and spin-orbit interaction yields good estimates for $E_0$ of both structural phases. Further analysis relates the optical gap with the crystalline-field energy as well as activation energy for electrical conduction. This last property characterizes the amorphous phase, thereby furnishing a link between the crystalline field and the activation energy and ultimately with the resistance drift exponent. Providing optical means to quantify resistance drift in PCMs could circumvent the need for fabricating expensive devices and performing time consuming measurements.

Keywords: optical contrast, phase change, resistance drift, optical band gap, phase-change materials, dielectric function

*avecrux5@gmail.com

Crystals are remarkably robust and stable. By contrast crucial properties of materials in their amorphous phase undergo change. For instance, phase-change materials (PCMs) suffer resistance drift by which their resistance grows temporally, in some cases, doubling in hours.[1, 2] At constant temperature, the PCM resistance exhibits a temporal behavior typified by

$$R(t) = R(t_0)(t/t_0)^{v_R} \qquad (1)$$

$R(t_0)$ being resistance at time $t_0$ and $v_R$ the drift exponent. Resistance drift is hardly trivial since it undermines the stability of stored data in PCMs, challenging their potential for multi-bit data storage applications.[3] Understanding this phenomenon has implications for related problems as well, such as optical contrast, whereby the dielectric function of the crystalline phase is significantly larger than that of the amorphous, a contentious issue despite a half century of controversy.[4-7] This relationship between optical and conduction properties, arising seemingly from disparate origins and which we trace to a common source in this paper, offers insights and efficient avenues for diagnostics.

Resistance drift is known to exist in spin glasses, disordered superconductors, granular materials, colloids and similar systems and even, analogically, plant roots.[8-13] For PCMs, various explanations have been offered: relaxation of bonds and vacancies, release of compressive stress, structural relaxation of amorphous PCM, binding energy changes brought on by relaxation, band gap widening upon annealing.[14-17, 2] One gleans from these that resistance drift originates from structural relaxation in the material. This is true, but without denying it, one observes that structural change cannot be the *direct* and *immediate* cause of resistance drift, in much the same way that breaking of hydrogen bonds in ice is not the direct cause of melting since the defining change of ice into water remains obscure.[18] We deal with resistance drift in this spirit: optical contrast provides 'smoking-gun' proof that PCM has amorphized.

We focus on PCMs GeTe, $Sb_2Te_3$ and $Bi_2Te_3$ whose ground state configurations are Ge: [Ar] $3d^{10}4s^24p^2$; Te: [Kr] $4d^{10}5s^25p^4$, Sb: [Kr] $4d^{10}5s^25p^3$; Bi: [Xe] $4f^{14}$ $5d^{10}$ $6s^2$ $6p^3$. Their valence bands (vb) are dominantly Te $5p^4$ orbitals and their conduction bands (cb) primarily Ge $4p^2$, Sb $5p^3$ and Bi $6p^3$ orbitals.[19, 20, 7, 21] Fortuitously, their crystalline and amorphous forms share like dynamics around the Fermi energy. Crystalline and amorphous GeTe, $Sb_2Te_3$ and $Bi_2Te_3$ have similar density of states (DOS) plots near the Fermi point.[22, 23] *a*-GeTe displays bonding angles close to 90° and is a lone-pair semiconductor.[24] A comparable configuration for $Sb_2Te_3$ is expected from currently available evidence [25]. There is, however, ample evidence of chiefly 90° bonding networks in $Sb_2Te_3$.[26, 27] *Ab initio* studies by Guo et al confirm that *a*-$Bi_2Te_3$ and *a*-$Sb_2Te_3$ show strikingly matching electronic properties.[23] A model of the valence (vb) and conduction bands (cb) first introduced by Mott and Davis and amplified by others will frame our discussion (see **Figures 1a** and **b**).[28-29] Following a model of topological insulators, the cb and vb of the crystalline phase are assumed to have opposite parities, in virtue of inversion symmetry.[30, 31] The lone pair (lp) electrons turn out to be $p_z$ orbitals.

Following studies of polymer films, there is consensus that amorphous materials, including PCMs, relax toward an energetically favorable ideal glass state, proceeding through transitions across more stable intermediate states.[32] Such transitions are described by an activation energy $E_A$ (or several) related with structural relaxation, which may be time and temperature dependent.[16] This leads to a quantitative picture of the drift phenomenon.[16, 17] By showing that the electron tunnelling activation energy in PCMs, which is of electronic origin, can be extracted from the dielectric function, we link the optical contrast to the resistance drift problem. Importantly, this may provide an optical means to quantify resistance drift in PCMs without need for fabricating expensive devices and performing time consuming measurements.

In this paper, we consider first the formation of the energy dispersion close to the Fermi point. Then we show how this picture changes when parameters of the original system undergo adiabatic change because of amorphization. We take the view that amorphization is a process in time by which a PCM crystal takes up its amorphous phase (or vice versa). Although this process necessarily involves structural changes, our treatment focusses on the optical contrast problem and its link to resistance drift. After linking with the dielectric function, we establish a connection with the optical band gap and finally with resistance drift.

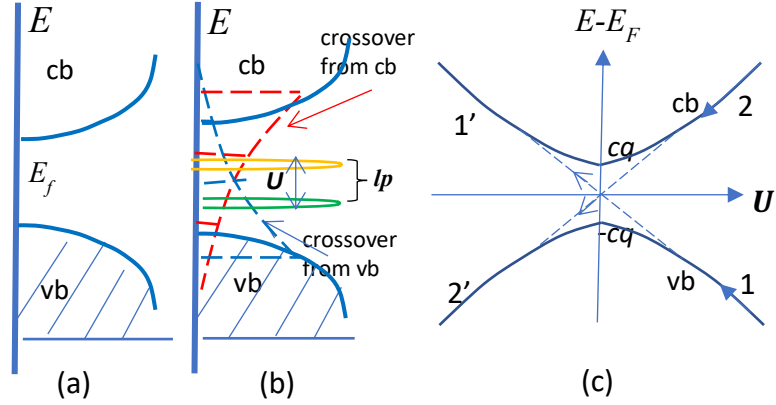

Figure 1. Schematic DOS for a) crystalline and b) amorphous PCM displaying the cb and vb. b) Localized band tail states and non-bonding lone-pair states occur near midgap. c) Pairs of electrons of opposite spins from the cb and vb couple through Friedel oscillations and open a gap. See text for discussion.

*Friedel oscillations:* Defects play a crucial role in PCMs.[28-29] A charged impurity in a metal induces an exponentially decaying electric screening, following the Debye-Hückel description.[33] Due to the quantum nature of electrons as waves obeying Fermi statistics, the interaction energy has a singularity at wave number $2k_0$ (the diameter of the Fermi surface), leading to a long-range algebraically decaying interference pattern of the electron density. According to Friedel, near interfaces, near defects, edges or impurities these oscillations vary significantly from those of the bulk electron gas.[34, 35] Moreover, electrons propagating in opposite directions close to the Fermi energy $E_F$ couple with each other through the charge-density oscillations thereby opening a gap.[36] In metals, this effect is overshadowed by large charge densities therein and is not easily observed. In amorphous PCMs, with their reduced charge concentrations, these oscillations face weaker competition; moreover, the role of impurities in metals can be taken by defects responsible for loss of long-range order on amorphization.[29] Nevertheless, it should be noted that Friedel oscillations have been studied even in connection with noninteracting electrons.[34b] In the context of PCM and dichalcogenides, Friedel oscillations have been of interest recently.[37] Because we are not trying to detect Friedel oscillations, but rather use them in the wider framework of studying the crystalline to amorphous transition in PCM and vice versa, it is not necessary to delve into finer details since we know these oscillations exist in PCM.

To describe this quantitatively, we assume low electron density, each crystal lattice site being occupied by $n = n_\uparrow + n_\downarrow \ll 1$ electrons (↑, ↓ indicating spin). An itinerant electron contributes an amount $\delta n \ll n$ to the lattice-site charge. In the narrow-energy-band picture, the Coulomb energy originating from a pair of electrons of opposite spins, occupying the same site, contributes $U(n_\uparrow + \delta n_\uparrow)(n_\downarrow + \delta n_\downarrow) - U n_\uparrow n_\downarrow \approx \frac{1}{2} U n \delta n$, $U=$ Hubbard energy.[29] See Figure 1c. Let $\langle n \rangle = \mathcal{V} \rho(x)$, $\mathcal{V} =$ unit-cell volume, then $\rho(x) = \rho_0 + \rho_1 \cos(Qx + \phi)$, $\rho_0$ being the charge density, $\rho_1$ the amplitude of charge oscillations, and $Q$ the wave vector. The Friedel Coulomb energy is

$$H_1 = \frac{1}{2} U \mathcal{V} \rho_1 \sum_i c_i^\dagger c_i \cos(Qx_i + \phi), \tag{2}$$

the $c_i$'s being quantized operators satisfying $\langle c_i^\dagger c_i \rangle = \delta n$. Consider two states which, in the absence of $H_1$, are plane waves $|k_x = Q/2 + q\rangle = e^{i(Q/2+q)x}/\sqrt{\mathcal{V}}$ and $|k_x = -Q/2 + q\rangle = e^{-i(Q/2-q)x}/\sqrt{\mathcal{V}}$, with energies $E(Q/2 \pm q) = E(Q/2) \pm \hbar c q$, $c =$ group velocity. Their matrix element is $\frac{1}{2} U \mathcal{V} \rho_1 \langle -Q/2 + q | \cos(Qx_i + \phi) | Q/2 + q \rangle = \frac{1}{4} U \rho_1$ and the perturbation matrix $\begin{pmatrix} E - E(Q/2) + \hbar c q & U\rho_1/4 \\ U\rho_1/4 & E - E(Q/2) - \hbar c q \end{pmatrix}$ yields the dispersion[37]

$$E = E(Q/2) \pm \sqrt{(\hbar c q)^2 + (U\rho_1/4)^2}, \tag{3}$$

Hence a gap opens at $E(Q/2)$. At equilibrium, $Q = 2k_F$, and the gap is at $E_F$. It arises from electron-electron interaction and is not an artifact of the Peierls distortion.[36] The model is structurally non-specific and can apply to amorphous structures with structural defects.

*Landau-Zener crossover*: Consider next amorphous PCMs which undergo temporal change. This property can be accommodated within the above considerations by allowing some parameters of the system to undergo adiabatic change over time as the material transitions from its crystalline phase to the amorphous. By *adiabatic* we mean it is slow change of a material relative to other atomic transitions. It is at the periphery between statics and dynamics. Anderson had advocated the concept of negative-'$U$' (negative Hubbard) energy by which two electrons situated at the same center attract in spite of their Coulomb repulsion.[29b] This would explain double occupancy of a localized state becoming more favorable than single occupancy of two localized states.[29] A mechanism that can give rise to negative-'$U$' is the inclusion of electron-lattice interactions. Taking our cue from resistance drift, we assume that as the material is undergoing phase change, the Hubbard energy undergoes adiabatic linear change spanning the range of positive and negative values, and cast the perturbation matrix as

$$H_1(t) = \begin{pmatrix} \gamma & \alpha t \\ \alpha t & -\gamma \end{pmatrix} \tag{4}$$

where energy is referred to $E_F$, $\alpha \ll 1$ is the Hubbard-energy rate of change and $\gamma = \hbar c q$. Note that while the material is crystalline, as in the case of Eq. (2), $\alpha t$ is just the constant $\frac{1}{4} U \rho_1$; but as the material transitions to the amorphous, the time dependence takes over. In the region of positive (negative) Hubbard energy, the material is crystalline (amorphous). We will note below that a general theorem (Dykhne formula) guarantees that the exact details (about $\alpha t$) of the time change do not alter the final outcome of our result. (Thus, we leave $\alpha$ unspecified.) Undoubtedly structural change precedes and accompanies this change, but we do not discuss it. As an aside, a logarithmic time dependence had been proposed by LeGallo.[16] Also pressure-induced shift from positive to negative $U$ has been noted in $c$-Si.[38] Through a unitary transformation, $H_1$ is recast as $\begin{pmatrix} -\alpha t & \gamma \\ \gamma & \alpha t \end{pmatrix}$, which is the standard form employed in Landau-Zener (LZ) calculations.[39, 40] Unlike Eq. (2), the charge-density amplitude is no longer constant, and we expect population transfer between levels (3), when swept past an avoided crossing.[40]

The (amorphous) Schrodinger equation is $i\hbar \dot\psi = \begin{pmatrix} -\alpha t & \gamma \\ \gamma & \alpha t \end{pmatrix}\psi, \psi = \begin{pmatrix} C_1(t) \\ C_2(t) \end{pmatrix}$. For the time-dependent problem, we choose boundary conditions, $C_2(-\infty) = 0$, $|C_1(-\infty)| = 1$. That is, the system is initially prepared in state $C_1$. The solutions are $C_1 = -\frac{\hbar\xi}{\gamma} A_- D_{-i\Delta}(-iz)$, $C_2 = A_- D_{-i\Delta-1}(-iz)$, where $\Delta = \frac{\gamma^2}{2\hbar\alpha}, \frac{\alpha^2}{\hbar^2 \xi^4} = -\frac{1}{4}$, $z = \xi t$, $D$'s are Weber functions and $A_-$ a normalization constant. See refs 40, 41 for calculations. The probability that the system remains in state $C_1$ after a long time is the Landau-Zener (LZ) result, $e^{-2\pi\Delta}$. At $t = 0$, it is $|C_1(0)|^2 = \frac{1}{2}(1 + e^{-\pi\Delta})$. The LZ probability can be derived *without* solving Schrodinger's equation by using Dykhne's formula.[42] This is important because Dykhne's formula only requires that the perturbing potential change adiabatically so its validity is stronger than LZ's result.[43] LZ's result predicts that for two time-dependent adiabatic states related by the time-dependent Schrodinger equation, an initially localized state evolves and splits into both states at the crossing. We estimate $\Delta \sim 10^2$, so $|C_1(0)|^2 \cong \frac{1}{2}$.[41] Thus, the crossing of levels occurs for any reasonable interaction provided the process takes place adiabatically.

To return to PCMs, first regard the energies (3) as corresponding to the top (bottom) of the vb (cb) in the *crystalline* phase, the quantity $cq > 0$ parametrizing the energies (Figure 1c). For the crystalline case, $\frac{1}{4}U\rho_1$ was constant. In the amorphous case, $\alpha t$ is the *time dependent* matrix element. A crossover (indicated by dashed lines) means that, in time, electrons near the *top* of the vb now occupy the *bottom* of the cb. Similarly, electrons close to the bottom of the cb now occupy the top of the vb. (This crossover leads to changes in the amorphous phase which explain the optical contrast.) Because $|C_1(t)|^2 < 1$, the crossover does not imply that the vb and cb have exchanged places. A partial exchange has taken place: these are the $p_z$ orbitals.

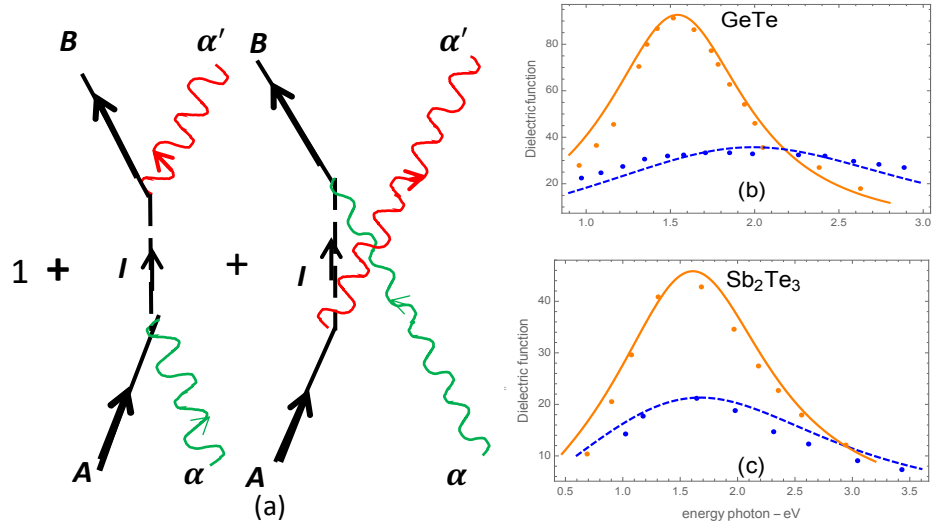

Figure2. a) Zeroth and second order space-time photon-electron diagrams. b) & c) $\varepsilon_2$ for crystalline (orange) and amorphous (blue) GeTe and Sb$_2$Te$_3$. WDD parameters for GeTe: crystalline, $E_d = 90$ eV, $E_0 = 1.6$ eV, $\hbar\gamma = 1.0$ eV; amorphous, $E_d = 80$ eV, $E_0 = 2.3$ eV, $\hbar\gamma = 2.4$ eV. For Sb$_2$Te$_3$, crystalline, $E_d = 70$ eV, $E_0 = 1.7$ eV, $\hbar\gamma = 1.4$ eV; amorphous, $E_d = 56$ eV, $E_0 = 2.4$ eV, $\hbar\gamma = 3.2$ eV. Data for: Sb$_2$Te$_3$ from ref. 44; GeTe from ref. 45.

*Dielectric function:* We extract experimental input from the dielectric function $\varepsilon(\omega)$. It is related to the zeroth and second order Feynman diagrams (**Figure 2a**).[7,41] From the theory of light,[46, 41, 47] the scattering amplitude is

$$f(p) \propto 1 - \frac{1}{m}\left[\frac{\langle B|\boldsymbol{p}\cdot\boldsymbol{\varepsilon}^{(\alpha')}|I\rangle\langle I|\boldsymbol{p}\cdot\boldsymbol{\varepsilon}^{(\alpha)}|A\rangle}{E_I-E_A-\hbar\omega} + \frac{\langle B|\boldsymbol{p}\cdot\boldsymbol{\varepsilon}^{(\alpha)}|I\rangle\langle I|\boldsymbol{p}\cdot\boldsymbol{\varepsilon}^{(\alpha')}|A\rangle}{E_I-E_A+\hbar\omega'}\right] \quad (5)$$

This relates with the Wemple-DiDomenico (WDD) parametrization, $\varepsilon(\omega) = \varepsilon_1 + i\varepsilon_2 = 1 + \frac{E_d E_0}{E_0^2 - (\hbar\omega)^2 - i\hbar\omega\hbar\gamma}$, $E_0$ being the oscillator energy; $E_d$, the dispersion energy and $\hbar\gamma$, the damping energy (which we included by hand).[48] These yield $E_0 = E_I - E_A = \hbar\omega_{IA}$ and $E_d = 2m\omega_{IA}^2|\langle I|\boldsymbol{x}^{(\alpha)}|A\rangle|^2$. $E_0$ is the intermediate state (*I*) energy while $E_d$ the photon absorption/emission coupling.

Compared with its crystalline counterpart, the amorphous absorption $\varepsilon_2$ is blue shifted, diminished and broadened (Figures 2b, c). We explain these distinguishing features. For chalcogenides the p-electrons experience the field from the ions framing their environment *and* the spin-orbit interaction (SOI), which is strong because of the large atomic number. The oscillator energy originates from the electrons' interaction with the crystalline field $E_{crys}$ and SOI. Because the amorphous phase is unstable relative to its counterpart, the crystalline phase should have smaller $E_{crys}$ than the amorphous. Since $E_0$ is nearly linear in $E_{crys}$ (see below), $E_0$ is therefore larger (blue shifted) for the amorphous phase.

Next consider the formula above for $E_d$. The overlap $\langle I|\boldsymbol{x}^{(\alpha)}|A\rangle$ is greater in the crystalline case than in the amorphous because of long range order. This stronger crystalline response to light explains its larger dispersion energy $E_d$. Finally, when the incoming photon energy equals the oscillator energy, $E = E_0$, the corresponding absorption is $\varepsilon_2(E_0) \approx E_d/\hbar\gamma$ (see WDD). With lower $E_d$ for the amorphous phase, $\varepsilon_2(E_0)$ should also be lower. However, it generally happens that $\varepsilon_2(E_0)$ is even smaller, which is explained by a larger damping $\hbar\gamma$ for the amorphous phase, hence broadening of its spectrum.

*Model for the oscillator energy:* Our model for $E_0$ was described elsewhere but we summarize it for completeness.[7] As noted above, the shift of $E_0$ during amorphization explains the blue shift of $\varepsilon_2$. It is a crucial quantity. The SOI is given by $H_{so} = \zeta\boldsymbol{\ell}\cdot\boldsymbol{s}$, $\boldsymbol{\ell}(\boldsymbol{s})$ being the orbital angular momentum (spin) and $\zeta$ the spin-orbit strength. The $p_x$ and $p_y$ orbitals define the atomic-layer plane, while $p_z$ orbitals are perpendicular to it.[31, 49] Thus, the planar $p_x$, $p_y$ orbitals experience the same crystalline field $E_{p_x} = E_{p_y}$, which we call $E_{CF}$. For convenience, $E_{crys}$ for $p_z$ orbitals is set to 0 when the SOI is switched off.[49] We treat $E_{crys}$ and SOI as perturbations on the $p$ orbitals. For the basis set we choose standard $p$-orbitals $\{p_x\uparrow, p_x\downarrow, p_y\uparrow, p_y\downarrow, p_z\uparrow, p_z\downarrow\}$, $p_i$'s being the radial part of the electron wave function multiplied by a real spherical harmonic. The Hamiltonian, in a one-atom system with valence p orbitals, is

$$H = \begin{pmatrix} E_{CF} & 0 & \frac{i\zeta}{2} & 0 & 0 & \frac{\zeta}{2} \\ 0 & E_{CF} & 0 & \frac{-i\zeta}{2} & \frac{-\zeta}{2} & 0 \\ \frac{-i\zeta}{2} & 0 & E_{CF} & 0 & 0 & \frac{i\zeta}{2} \\ 0 & \frac{i\zeta}{2} & 0 & E_{CF} & \frac{i\zeta}{2} & 0 \\ 0 & \frac{-\zeta}{2} & 0 & \frac{-i\zeta}{2} & 0 & \frac{i}{2}0 \\ \frac{\zeta}{2} & 0 & \frac{-i\zeta}{2} & 0 & \frac{i}{2}0 & 0 \end{pmatrix} \quad (6)$$

For the vb (cb) the crystal field $E_{CF}$ takes negative (positive) values (assumed constant).[31, 48] Due to inversion symmetry, $\zeta$ takes on opposite signs for the valence and conduction electrons.[31, 50] From Eq. (6) we obtain the doubly degenerate SOI-induced energy splittings

$$E_1 = \frac{1}{2}(2E_{CF} - \zeta), j = \frac{3}{2}, |m_j| = \frac{3}{2}$$
$$E_{2(3)} = \frac{1}{4}\left(2E_{CF} + \zeta - (+)\sqrt{4E_{CF}^2 + 4E_{CF}\zeta + 9\zeta^2}\right), j = \frac{3}{2}(\frac{1}{2}), |m_j| = \frac{1}{2}$$ (7)

the total angular momenta $j = \frac{3}{2}, \frac{1}{2}$ and their z-projection $m_j$ serving as convenient labels.

How does amorphization alter this picture? The LZ mechanism implies 'crossing' between bands during amorphization, that is, the cb $\{p_z \uparrow, p_z \downarrow\}$ orbitals cross over to the vb and vice versa (Figure 1(c)). Not all orbitals cross over since the LZ probability is not unity. Orbitals that crossed over, *retain* their native parities. Therefore, in the amorphous case we find mixed parity states in the crossover bands; they are the non-bonding lone-pairs in Figure 1(b). Thus, some matrix elements involving the $p_z$ orbitals in Eq. (6) *change* sign, specifically the matrix elements on the fifth row and on the sixth column. On introducing these changes in Eq. (6) the new solutions are seen to depart from the simple picture of Eq. (7), which has solutions for *all* values of $E_{\text{crys}}$. For the amorphous case, a band gap develops since real solutions exist only *beyond* a critical crystalline field (see below). This critical field energy is identified with the amorphous band gap. (Alternatively, the sign changes could be on the sixth row and fifth column. But the effect is identical to the one just described, because this scenario and the previous, i.e., for the fifth row and sixth column, are Hermitian conjugates of each other.) The sign change is ultimately responsible for the blue shift of $E_0$ and explains the optical contrast of PCM.

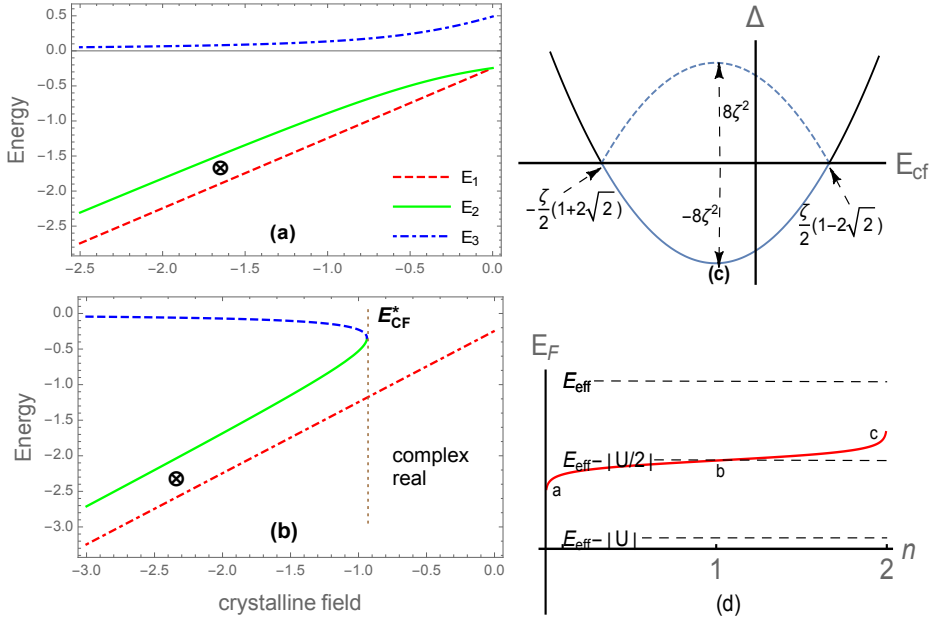

Figure 3. Interplay between $E_{CF}$ and SOI on vb p-orbitals. a) Crystalline field only applies to $p_x$-$p_y$ orbitals. ⊗ indicates the average energy of the four Te $p$ orbitals when $E_{CF}$ = -1.65 eV. b) Amorphous case. Solutions $E_{2,3}$ are complex in general. Beyond $E_{CF}^* = -0.93$ eV, $E_{2,3}$ are real. We identify $E_{CF}^*$ with the optical gap. ⊗ indicates the average energy of the Te $p$ orbitals when $E_{CF} \approx$ -2.4 eV. c) Plot of $\Delta$ vs $E_{cf}$. Dashed parabola is the inverted image of $\Delta$ when it is negative. d) Fermi energy $E_F$ as a function of occupancy $n$ for negative $U$.

Results for the crystalline phase are displayed in **Figure 3a**, where $\zeta = 0.49$eV for Te. [51] When $E_{CF} = 0$, solutions $E_{1,2}$ are degenerate; the SOI splits the $j = \frac{1}{2}$ doublet and the $j = \frac{3}{2}$ quadruplet. As $E_{CF}$ grows, the quadruplet splits further into two doublets. There being four Te valence electrons, these follow the pair of lower doublet solutions. $Sb_2Te_3$ and $Bi_2Se_3$ have similar lattice parameters and we expect both to have similar $E_{CF}$, which is ~-1.3 eV for $Bi_2Se_3$. [49] Using -1.65 eV, we locate $E_0$ at the point denoted by ⊗, i.e. $E_0 \approx 1.7$ eV, identical with the value quoted in Figure 2c.

Figure 3(b) concerns the amorphous case. Solutions are now complex because the discriminant Δ can be negative (see next paragraph). However, beyond the point labelled $E_{CF}^*$, all solutions are real, i.e., left of $E_{CF}^*$, *all* eigenvalues are real. For $Sb_2Te_3$, $E_{CF}^* = -(1 + 2\sqrt{2})\frac{\zeta}{2} \approx -0.94$ eV. We identify this with the band gap $E_g$. Note the shift of $E_0$ to $\approx -2.4$eV. Observe that $E_{1,2}$ are almost linear in $E_{CF}$.

*Crystalline energy:* We examine closely Eq. (7), which applies to the crystalline case. For the amorphous, it still holds but the discriminant inside the square root is now $\Delta = 4E_{cf}^2 + 4E_{cf}\zeta + \zeta^2 + 8\zeta\xi$ where $\xi(= -\zeta)$ is the SOI constant whose sign changed during amorphization (due to crossover). Δ vanishes when $2E_{cf} + \zeta = -2\sqrt{2}\zeta$ and, in the field of negative $E_{cf}$ (corresponding to the vb), $E_{cf} = -\frac{1}{2}\zeta(1 + 2\sqrt{2}) \approx -0.94$eV for the threshold field which we identified with the *amorphous-phase* gap, i.e., $|E_g| \approx 0.94$eV. For the *crystalline* phase, $\xi = +\zeta$, so there was *no* sign change: $\Delta > 0$. (Note: $E_{CF}$ denotes the crystal field for the crystalline phase while $E_{cf}$ the crystal field for the amorphous.) We have from Eq. (7), $E_1 = \frac{1}{2}(2E_{CF} - \zeta)$. Solution $E_2$ is close by as Fig. 3a shows. With $E_1$ representing both solutions approximately, we identify $E_1$ with the *crystalline* oscillator energy $E_0$ for the following crystal field, $E_{CF} = E_g - \zeta\sqrt{2} = -\frac{1}{2}\zeta(1 + 4\sqrt{2})$. For *this* choice of $E_{CF}$, we find $E_0 = |-\zeta(1 + 2\sqrt{2})| = 2E_g$. The observation that $E_0 \sim 2E_g$ was made by Tanaka 40 years ago [52]. Table I, columns 2 and 3 demonstrate that Tanaka's observation applies to technologically important binary and ternary chalcogenides. We provided a physical basis for this observation.

Table I: Oscillator energies ($E_0$), optical band gaps ($E_g$)
& activation energies ($E_A$) in eV for some chalcogenides.

|  | $E_0$ | $E_g$ | $E_A$ |
|---|---|---|---|
| $Sb_2Te_3$ | 1.7[a] | 0.8[e] | 0.28[k] |
| GeTe | 1.6[a] | 0.75[f] | 0.36[l] |
| $Bi_2Te_3$ | 1.0[b] | 0.43[g] | 0.34[m] |
| $Ge_2Sb_2Te_5$ | 1.9[c] | 1.0[h] | 0.42[n] |
| $Bi_2S_3$ | 3.4[d] | 1.9[i] | 1.0[o] |
| Se | 3.9[c] | 1.9[j] | 1.0[p] |
| $Ga_2Se_3$ | 4.09[q] | 2.05[r] | 0.9[s] |

[a]Fig. 2  [d]Ref. 54  [g]Ref. 57  [j]Ref. 60  [m]Ref. 63  [p]Ref. 66  [s]Ref. 69
[b]Ref. 53  [e]Ref. 55  [h]Ref. 58  [k]Ref. 61  [n]Ref. 64  [q]Ref. 67
[c]Ref. 41  [f]Ref. 56  [i]Ref. 59  [l]Ref. 62  [o]Ref. 65  [r]Ref. 68

Next we derive the activation energy $E_A$ for electron tunneling. In Figure 3c the (amorphous phase) discriminant Δ is plotted against $E_{cf}$. The portion below the $E_{cf}$ axis corresponds to $\Delta < 0$. No real solution for $E_{2(3)}$ exists there. The bottom of the parabola is $-8\zeta^2$ deep. In field theory, one makes a Wick rotation which inverts this 'potential.'[70] This is the dashed parabola whose height is $+8\zeta^2$. From $E_2$ and using *this* value for Δ, we obtain $E_2 = \frac{1}{4}(2E_{cf} + \zeta +$ $\sqrt{\Delta}) = \frac{1}{4}(2E_{cf} + \zeta + 2\sqrt{2}\zeta)$. The quantity $\frac{1}{4}(\zeta + 2\sqrt{2}\zeta) = E_g/2$ represents the barrier height as the electron tunnels to a new minimum, i.e., activation energy.[71] It is customary to represent semiconductor resistance by $R = R^* e^{E_A/k_B T}$.[17] The relation $E_A = E_g/2$ is also invoked but appears to be empirical.[16, 72, 73] Our discussion provides a framework for deriving it. See Table I. Going further, we noted that the crystalline field in the crystalline phase was $E_{CF} = E_g - \zeta\sqrt{2}$. Increasing this by another $\zeta\sqrt{2}$ gives the crystalline field for the *amorphous* phase, i.e. $E_{cf} = E_{CF} - \zeta\sqrt{2}$, which one can check from Figure 3b (i.e., point ⊗ in Fig. 3(a) is shifted by $\zeta\sqrt{2}$ in Figure 3(b)). Thus, the gap $E_g$, the crystalline field for the *crystalline* phase $E_{CF}$ and the crystalline field for the *amorphous* phase $E_{cf}$ are successively separated by $\zeta\sqrt{2}$.

*Resistance Drift:* Finally, we connect the optical gap with resistance drift. A defect $D$ may be occupied by two electrons. It is usually neutral ($D^0$) when singly occupied; charged +e ($D^+$) when unoccupied; and charged –e when doubly occupied ($D^-$). Under charge neutrality, the Fermi energy is midway between the singly and doubly occupied states. For $U < 0$, the $D^0$ state is energetically higher than both $D^+$ and $D^-$ states so equilibrium favors equal concentrations of $D^{+(-)}$ with almost none of $D^0$. As defect occupancy $n$ grows ($0 \leq n \leq 2$), the Fermi energy $E_F$ departs little from its position: it is pinned. Pinning is encapsulated in the solution derived by Adler and Yoffa:[74]

$$E_F \approx E_{\text{eff}} - \tfrac{1}{2}|U| + \tfrac{1}{2}k_B T \log\frac{n}{2-n}, \tag{8}$$

$E_{\text{eff}}$ being the bare defect energy. In the region labelled $c$ in Figure 3d, $n \approx 2$; let $n = 2 - \delta, \delta \ll 1$, and identify $\tfrac{1}{2}|U| + \tfrac{1}{2}k_B T \log 2$ with half the optical gap *and* $-\tfrac{1}{2}k_B T \log \delta$ with its time-dependent correction. We propose the following form, $\delta \sim t^{-\alpha}, \alpha = 1 - \frac{T}{T_0}$, where $T$ ($T_0$) is the operating (transition) temperature and $t \gg 1$ is time in reduced units, i.e. it is dimensionless. In arriving at this form, we assume a) linear response theory; b) adiabaticity; c) that the material ceases being amorphous on passing the transition temperature; d) absence of other thermal effects. Then the form for $\delta$ is unique. It should be noted that a corrected version of Eq (1) has been discussed by Wimmer et al which is not taken up in this paper.[83] Then resistance is given by $R = t^{\frac{1-\alpha}{4}}(R^* e^{E_g/2k_B T})$ so $\nu_R = (1 - \alpha)/4$ is the exponent of Eq. (1). Below we give in Table II a comparison between the calculated and observed values of $\nu_R$ for some chalcogenides. In the region labelled *b* in Figure 3d) we expect negligible drift.

Table II Transition temperatures and resistance drift exponents

|  | $T_0$ (K) | $T$ (K) | $\nu_R$ | expt value |
|---|---|---|---|---|
| $Ge_2Sb_2Te_5$ | 420[a] | 300 | 0.1 | 0.1[b] |
| GeTe | 508[c] | 323 | 0.09 | 0.12[d] |
| $Sb_2Te$ | 417[e] | 323 | 0.057 | 0.054[f] |
| $Ge_{15}Te_{85}$ | 493[g] | 323 | 0.086 | 0.05[f] |
| $Ge_{15}Sb_{85}$ | 604[h] | 323 | 0.12 | 0.13[f] |
| AIST $Ag_4In_3Sb_{67}Te_{26}$ | 451[i] | 323 | 0.07 | 0.06[f] |

[a]Ref. 75  [b]Ref. 1  [c]Ref. 76  [d]Ref. 77  [e]Ref. 78
[f]Ref. 79  [g]Ref. 80  [h]Ref. 81  [i]Ref. 82

In summary, we first recalled that Friedel oscillations couple electrons moving in opposite directions and supply an additional Coulomb energy to PCMs. When the material amorphizes this energy becomes time dependent and the Landau-Zener mechanism operates, leading to population transfer in both directions between the vb and the cb. Comparing the dielectric function with a theoretical model, we noted that the oscillator energy dominated the optical spectrum. A $k \cdot p$ calculation involving crystalline field and SOI yielded good estimates for $E_0$ for both amorphous and crystalline phases. We showed that $E_0$, the optical gap, the crystalline energies and activation energy are related in a simple way. This provided a theoretically grounded link from the optical gap to the electron tunnelling activation energy and concomitantly to the resistance drift exponent.

During the review process we have become aware of other recent publications that link resistance drift as having electronic rather than structural origin through lone pairs.[84]

We gratefully acknowledge support from the Singapore Ministry of Education (project number MoE 2017-T2-1-161). This work was conducted under the auspices of the SUTD-MIT International Design Center (IDC).

Supporting Information: Supporting information is available from the Wiley Online Library or from the author.